\documentclass[%
 reprint,
 amsmath,amssymb,
 aps,
]{revtex4-2}

\usepackage[T1]{fontenc}
\usepackage[latin9]{inputenc}
\usepackage{amsbsy}
\usepackage{amstext}
\usepackage{amssymb}
\usepackage{graphicx}
\usepackage{dcolumn}
\usepackage{bm}

\usepackage[unicode=true,
bookmarks=true,bookmarksnumbered=true,bookmarksopen=true,bookmarksopenlevel=1,
breaklinks=false,pdfborder={0 0 0},pdfborderstyle={},backref=false,colorlinks=false]
{hyperref}
\hypersetup{pdftitle={Your Title},
	pdfauthor={Your Name},
	pdfpagelayout=OneColumn, pdfnewwindow=true, pdfstartview=XYZ, plainpages=false}

\usepackage{fancyhdr}

\makeatletter

\begin{document}

\preprint{APS/123-QED}

\title{Exceptional Point Degeneracy as Desirable Operation Point of Oscillator Array with Discrete Nonlinear Gain and Radiating Elements}

\author{Alireza Nikzamir}
\author{Filippo Capolino}%
 \email{f.capolino@uci.edu}
\affiliation{Department	of Electrical Engineering and Computer Science, University of California, Irvine, CA 92697 USA}

\begin{abstract}
	An oscillator array prefers to operate at an exceptional point of degeneracy (EPD) occurring in a waveguide periodically loaded with discrete nonlinear gain and radiating elements. The system maintains a steady-state degenerate mode of oscillation at a frequency of 3 GHz, even when the small-signal nonlinear gain values are nonuniform along the array. Contrarily to the original expectation of zero phase shift associated to the designed EPD using small-signal gain, after reaching saturation, the time domain signal in consecutive unit cells displays a $\pi$ phase shift. Hence, we demonstrate that the saturated system oscillates at a distinct EPD, associated to a $\pi$ phase shift between consecutive cells, than the one at which the system was originally designed using small-signal gain. This new EPD at which the nonlinear system is landing is associated to higher power efficiency. Finally, we demonstrate that the oscillation frequency is independent of the length of the array, contrarily to what happens ordinary oscillating systems based on one-dimensional cavity resonances. These findings may have a high impact on high-power radiating arrays with distributed active elements.

\end{abstract}

\maketitle

\thispagestyle{fancy}


\section{\label{sec:level1}Introduction}

Exceptional points of degeneracy (EPDs) in waveguides have become increasingly popular in the fields of electromagnetics, photonics, and radio frequency (RF) circuits \cite{Klaiman2008Visualization, Guo2009Observation, ruter2010observation,Wood2015Degenerate,Schnabel2017PT-symmetric,Abdelshafy2019Exceptional}. EPD is the condition at which two or more eigenmodes coalesce in their eigenvalues and eigenvectors \cite{Vishik_1960_Solution,Kato1966Perturbation,Lancaster1964On,Seyranian1993Sensitivity}. The term EP has been in use since \cite{Kato1966Perturbation}; as was also emphasized in \cite{Berry2004Physics}, the key physics feature is the "degeneracy", and that is the reason for the "D" in EPD. At the EPD, the system matrix or transfer matrix representing the mode evolution in the system is similar to a matrix containing a non-trivial Jordan block \cite{Figotin2005Gigantic, Bender2010PTsymmetry}. The order of the degeneracy is the number of coalescing eigenmodes at the EPD. Close to an EPD of order $2$ in a waveguide, the dispersion relation between frequency and wavenumber is $(\omega-\omega_{e})\propto(k-k_{e})^{2}$, where the subscript $e$ denotes EPD. When the system refractive index obeys $n(x)=n^{\ast}(-x)$,  where $x$ is a coordinate in the system orthogonal to the propagation direction $z$, and $*$ is the complex conjugate, the system is parity time (PT) symmetric  \cite{Bender1998Real,Barashenkov_2013,El_Ganainy2007Theory}. PT symmetry is a condition that enables the occurrence of EPDs with a degenerate real eigenvalue  \cite{Guo2009Observation,Schnabel2017PT-symmetric,heiss_2012physics,ruter2010observation}, though it is not a necessary condition to get EPDs \cite{Nada2017Theory,Mealy2020EPDinEbeam}. EPDs in periodic structures can be classified into two categories: (i) those obtained with gain and/or loss \cite{El_Ganainy2007Theory,Guo2009Observation,ruter2010observation,Klaiman2008Visualization, Yazdi2021Third, Mealy2020EPDinEbeam,tuxbury2022non}, and (ii) those obtained without gain and loss \cite{Figotin2005Gigantic, Gutman2012Bistability, Apaydin2012Experimental, Abdelshafy2019Exceptional, Nada2017Theory, Nada2021FrozenMode} (note that in some of these papers, the author did not use the term EP). The occurrence of an EPD enables special and unique physical features  that can be used in different applications from RF to optics. The EPD concept has been proposed to provide notable enhancements in the performance of oscillators and amplifiers: there are two main categories of these applications, classified according to the presence or absence of loss and gain. The first category involves EPD in waveguides without loss and gain, such as the DBE or SIP laser concepts \cite{Veysi2018Degenerate, Herrero23SIPlaser} that exhibit a new threshold scaling law, and in microstrip waveguides for arrayed antennas and providing a stable oscillation \cite{Abdelshafy2021Distributed}. The second category involves waveguides where EPDs are obtained thanks to the presence of loss and gain (PT symmetry is an example) that leads to the concepts of arrays of radiators with high-output power, backward-wave oscillators with distributed power extraction, etc \cite{Yazdi2021Third, Mealy2020EPDinEbeam, Mealy2021XbandBWO, Abdelshafy2021Exceptional} (these examples do not involve PT symmetry; one involves Glide-Time symmetry). In particular, accurate particle-in-cell simulations have shown that this kind of EPD in a waveguide with distributed gain and power extraction may enhance the performance of high-power electron beam devices \cite{Mealy2021XbandBWO,mealy2021high-Power}.

\begin{figure}[t]
		\begin{centering}
			\includegraphics[width=3.5in]{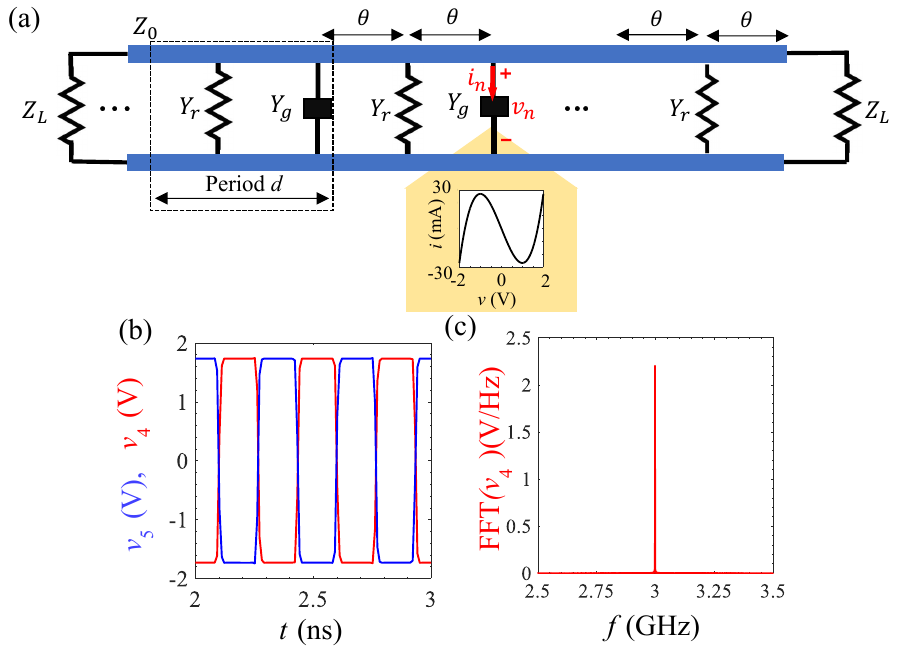}
			\par\end{centering}
		\caption{(a) Schematic of a periodically loaded waveguide represented by its equivalent transmission line (TL). Each unit cell is made of two TL segments with characteristic impedance $Z_{0}$ and same electrical lengths $\theta$, loaded with a lossy shunt element $Y_{r}$ representing a radiator (e.g., an antenna) and a shunt nonlinear gain element $Y_{g}$. (b) Time-domain voltage signals $v_{4}(t)$ and $v_{5}(t)$ are evaluated at the two middle unit cells' active elements, for an array with $N=8$. Radiation is  given by $Y_{r}Z_{0}=2.5$ and the nonlinear small-signal gain is $gZ_{0}=0.5$. (c) Frequency spectrum of the voltage $v_{4}(t)$ shows the oscillation frequency $f_{osc}=3\:\mathrm{GHz}$.\label{fig:Waveguide_nolinear}}
	\end{figure}
	
Oscillators play a vital role in microwave, THz, and optics applications. At RF, there is interest in making oscillators that offer stable oscillation frequency \cite{Frerking1978Crystal,Walls1986Measurements}, high-quality factor \cite{,An_Sun_Hyun1999K-band,Hosoya2000lowphase-noise}, loading independency \cite{Abdelshafy2021Distributed}, and high output power \cite{Kasagi2019Large-scale}. The EPD concept with gain and loss has been proposed to enhance the performance of distributed oscillators in various ways \cite{Abdelshafy2021Exceptional,Mealy2020EPDinEbeam,Mealy2021XbandBWO}. A method to design distributed oscillators with EPD is through the utilization of waveguide loaded with periodic gain and loss \cite{Abdelshafy2021Exceptional}. In this approach, losses represent the arrayed radiating elements, e.g., antennas.

	This paper presents a waveguide system modeled as a transmission line (TL), periodically loaded with discrete nonlinear gain and radiating elements as shown in Fig. \ref{fig:Waveguide_nolinear}(a). Rather than the general EPD conditions discussed already in \cite{Abdelshafy2021Exceptional}, here we explore the nonlinear features of the same structure, including the saturation from a nonuniform distribution of nonlinear gain elements. The analysis of nonuniform distribution of gain is extremely important in practice because it is impossible to guarantee that the active elements have the same value of gain along a waveguide, at both RF and optical frequencies. Variation in nonlinear gains over the arrayed structure, due to device tolerances,  may alter the oscillator operation and affect the power extraction from radiating elements. Therefore, we show that when utilizing the EPD concept in waveguide oscillators, even nonuniform distributions of nonlinear gain elements along the array lead to a stable oscillation regime, and this stationary regime leads to a uniform {\em saturated} gain distribution. To confirm the full degeneracy of the eigenmodes, we employ the "coalescence parameter" tool to demonstrate the coalescence of the eigenvectors. Also, we show that the system maintains a stable oscillation frequency even when varying the length of the structure (number of unit cells), and also when the nonlinear small-signal gains are not uniform along the array, and when loads on the two sides are varied. Finally, we discuss the degeneracy condition in the presence of an additional small reactance in parallel to the gain element, and identify the EPD using the coalescence parameter tool. We provide an example where a small capacitance is added to each gain element, and we demonstrate that the system exhibits stable oscillation at a different EPD. This approach shows that we can create a tunable oscillator by adding a small tunable capacitor in each unit cell. Additionally, we confirm that the system tends to oscillate at a state where the nonlinear saturated gain is diminished, bring the system to another EPD.
	
	\section{Oscillatory regime with nonlinear gain}

 	\begin{figure}[t]
		\begin{centering}
			\includegraphics[width=3.4in]{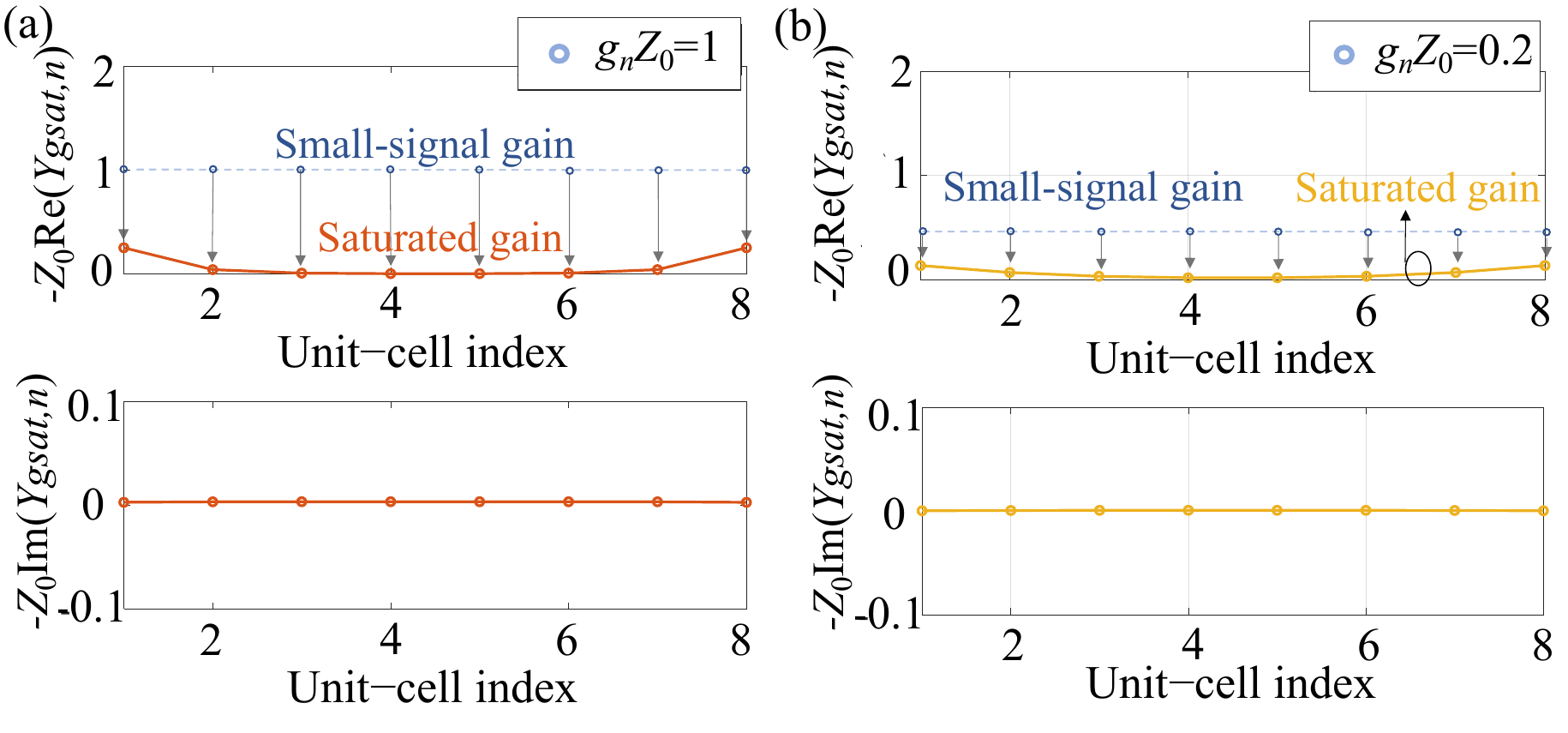}
			\par\end{centering}
		\caption{(a) Saturated gain calculated at each unit cell, found by Eq. \ref{eq:Effective_gain}, for two arrays with different nonlinear small-signal gain. Radiation losses are $Y_{r}Z_{0}=2.5$ in both cases. For two cases of arbitrary uniform nonlinear small-signal gain values $gZ_{0}=1$ (orange line) and $gZ_{0}=0.2$ (yellow line), the real and imaginary part of the \textit{saturated} gain will end up at $Y_{gsat,n}Z_{0}\approx0$ in each unit cell. \label{fig:Gain_variation}}
	\end{figure}
 
	The array oscillator consists of a waveguide, modeled as a TL with characteristic impedance $Z_{0}=50\:\Omega$, with a periodic distribution of $N$ lumped nonlinear gain elements, described by shunt admittances $Y_{g,n}$ with $n=1,2,...,N$ (not necessarily equal to each other), and $N+1$ radiating elements described by shunt admittances $Y_{r}$ (all equal to each other), both arranged periodically with period $d$ as shown in Fig. \ref{fig:Waveguide_nolinear}(a). This assumption reflects realistic scenarios where it is easy to make radiators that are very similar to each other, hence exhibiting the same admittance (e.g., dielectric resonator antennas, patch antennas, slot antennas, etc.) whereas it is almost impossible to ensure that the gain of each element is the same when using active components like transistors or even in the case of optically or electrically pumped lasers.
 
 For brevity, we only consider the case where radiating and gain lumped elements are separated by $d/2$. Therefore, each TL segment has electric length $\theta=k_{w}d/2$, where  $k_{w}=2\pi f/v_w$ is the wave propagation constant in the uniform TL segments where $f$ is the frequency, and the phase velocity $v_w$ is assumed to be dispersionless for simplicity. The array structure is symmetric, i.e., we terminate the left and right ends with loads $Z_{L}=Y_{r}/2$, and on the right side there is an extra shunt radiating element $Y_{r}$. Thus, the structure with $N$ unit cells has $N$ nonlinear gain elements, $N+1$ radiating elements, and left and right load terminations. Each gain element is described by a negative small-signal conductance $Y_{g,n}=-g_n$ and  saturation effect, where the current $i_{n}$ and voltage $v_{n}$, with $n=1,2,...,N$, are related by the $i-v$ cubic model 
	
	\begin{equation}
		i_n=-g_{n}v_n+\alpha_n v_n^{3},
        \label{eq:Nonlinear_gain}
	\end{equation}
	and $\alpha_n=g_{n}/3$  (unit of ${\rm S}/{\rm V}^2$) describes the saturation level. This cubic model provides a negative conductance $Y_{g,n}=-g_n$ for small voltage in the range between $-1\:\mathrm{V}<v_n<1\:\mathrm{V}$. Due to the 3rd-order nonlinearity, each active element saturates to an admittance value $Y_{gsat,n}$ that provides a gain $-{\mathrm{Re}}(Y_{gsat,n})$ to the system, which differs from the small-signal gain $g_{n}$. The saturated gain admittance (magnitude and phase) is found numerically by looking at the frequency components of the voltage and current, selecting the oscillation frequency $f_{osc}$, by using the fast Fourier Transform (FFT) as
	
	\begin{equation}
		\left\{ \begin{array}{c}
			\left|Y_{gsat,n}\right|=\frac{\left|\mathrm{FFT}\left(i_{n}\right)\right|_{f_{osc}}}{\left|\mathrm{FFT}\left(v_{n}\right)\right|_{f_{osc}}}\\
			\\
			\angle Y_{gsat,n}=\left.\left(\angle\mathrm{FFT}\left(i_{n}\right)-\angle\mathrm{FFT}\left(v_{n}\right)\right)\right|_{f_{osc}}
		\end{array}\right.,\label{eq:Effective_gain}
	\end{equation}
	where $\left|\,\right|$ represents the magnitude, and $\angle$ represents the phase. Numerical simulations are carried out in the time domain using the Keysight Advanced Design System (ADS) circuit simulator. 
 We first assume that the uniform (i.e., $g_n=g$, constant along the array) normalized nonlinear small-signal gain and radiating element are $gZ_{0}=0.5$ and $Y_{r}Z_{0}=2.5$, respectively. Figure. \ref{fig:Waveguide_nolinear}(b) shows the time domain oscillatory signals $v_{4}$ and $v_{5}$ in the middle of the array. Figure \ref{fig:Waveguide_nolinear}(c) shows the frequency spectrum of $v_{4}$ with the  fundamental frequency of oscillation of $3\:\mathrm{GHz}$, calculated by using the fast Fourier transform (FFT) of the saturated signal in the time window from $2\:\mathrm{\mu s}$ to $12\:\mathrm{\mu s}$, with $10^{6}$ points. We observe that the time domain signal at the two consecutive unit cells has a $\pi$ phase shift.
 
    To understand the observed oscillatory regime, we consider two other uniform cases with initial small-signal gain values of $gZ_{0}=1$ and $gZ_{0}=0.2$, and we calculate the cells' saturated gain using Eq. (\ref{eq:Effective_gain}) after reaching saturation.  Importantly, we observe that in both cases, after reaching saturation, the system  {\em still} oscillates at $f=3\:\mathrm{GHz}$, as observed in the previous case  with $gZ_0=0.5$.  Furthermore, we observe that the time domain voltages on the gain elements in consecutive unit cells still have a $\pi$ phase shift, as in the previous case with $gZ_0=0.5$. For the cases of $gZ_{0}=1$ and $gZ_{0}=0.2$, Figs. \ref{fig:Gain_variation}(a) and (b) show the real and imaginary parts of the saturated gain $-Y_{gsat,n}$ in each unit cell, which is mainly real positive. The real part $-\mathrm{Re}(Y_{gsat,n})$ of the saturated gain at steady state regime are smaller than $g$, in both cases. At this stage, it seems that the system tends to work at the point that has a saturated gain such that $-\mathrm{Re}(Y_{gsat,n}) << (1/Z_0)$. In the following, we analyze the modes of the structure to determine the characteristics of this specific point.

	\section{Formulation of a waveguide periodically loaded with discrete linear
		gain\label{sec:Formulation}}
	
	\begin{figure}[t]
		\begin{centering}
			\includegraphics[width=3.5in]{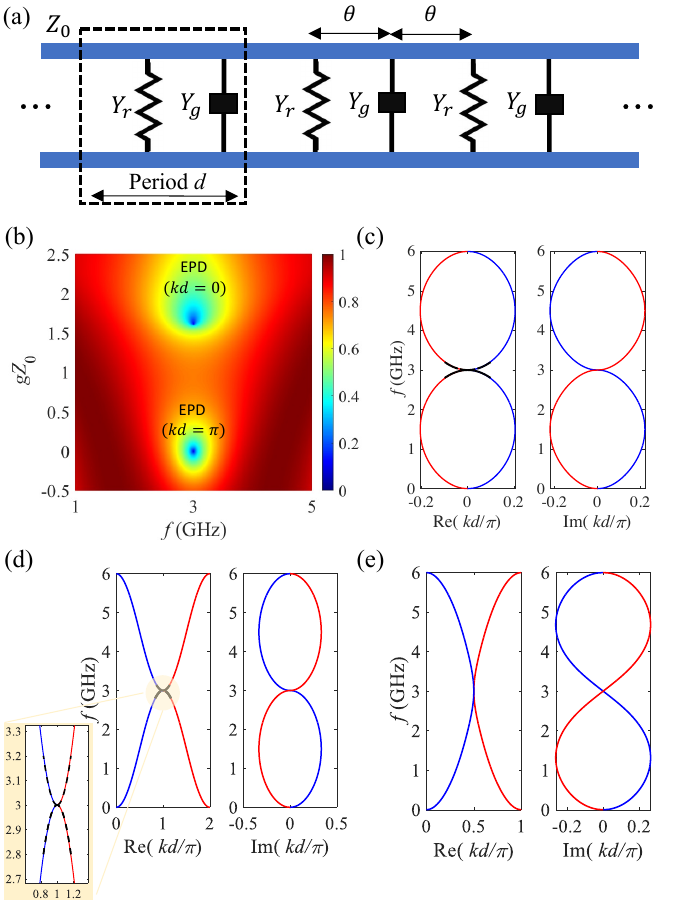}
			\par\end{centering}
		\caption{(a) Schematic of a periodic waveguide represented in terms of an equivalent TL with characteristic impedance $Z_{0}=50\:\Omega$, loaded periodically with a lumped loss $Y_{r}$ and linear gain $Y_{g}$ admittances. We assume $Y_{r}Z_{0}=2.5$ and $\theta=\pi/2$ at $3$ GHz. (b) The vanishing of the coalescence parameter shows two EPDs calculated from Eq. \ref{eq:Char_Eq}, for varying small-linear gain $g$. The two EPs are at $kd=0$ (for $gZ_{0}=1.6$) and $kd=\pi$ (for $gZ_{0}=0$). Dispersion	relation of the real and imaginary parts of the complex-valued wavenumber $k$ versus frequency for (c) $gZ_{0}=1.6$ and (d)  $gZ_{0}=0$. In the inset, the dispersion diagram is fit with the quadratic equation $(f-f_{e})=\pm\eta(k-k_{e})^{2}$ denoted by the black dashed line, with $\eta\approx7.153\times10^{4}\:\mathrm{m^{2}/s}$. (e) A case without supporting EPD ($gZ_{0}=0.8$). }
  \label{fig:2nd_EPD}
	\end{figure}
	
	Second order EPDs can occur in the waveguide under study. We find the system's eigenmodes by using the transfer matrix approach as in   \cite{Yazdi2021Third,Abdelshafy2021Exceptional,Buddhiraju2020Nonreciprocal}. We define the state-vector  ${\boldsymbol{\Psi}}(z)=[V(z),\:I(z)]^{\mathrm{T}}$ with  voltage $V(z)$ and current $I(z)$ along the waveguide,  and  $\mathrm{T}$ is the transpose action. The state vector in the periodic structure with a period of $d$ changes as 

  \begin{figure*}[t]
		\begin{centering}
			\includegraphics[width=7in]{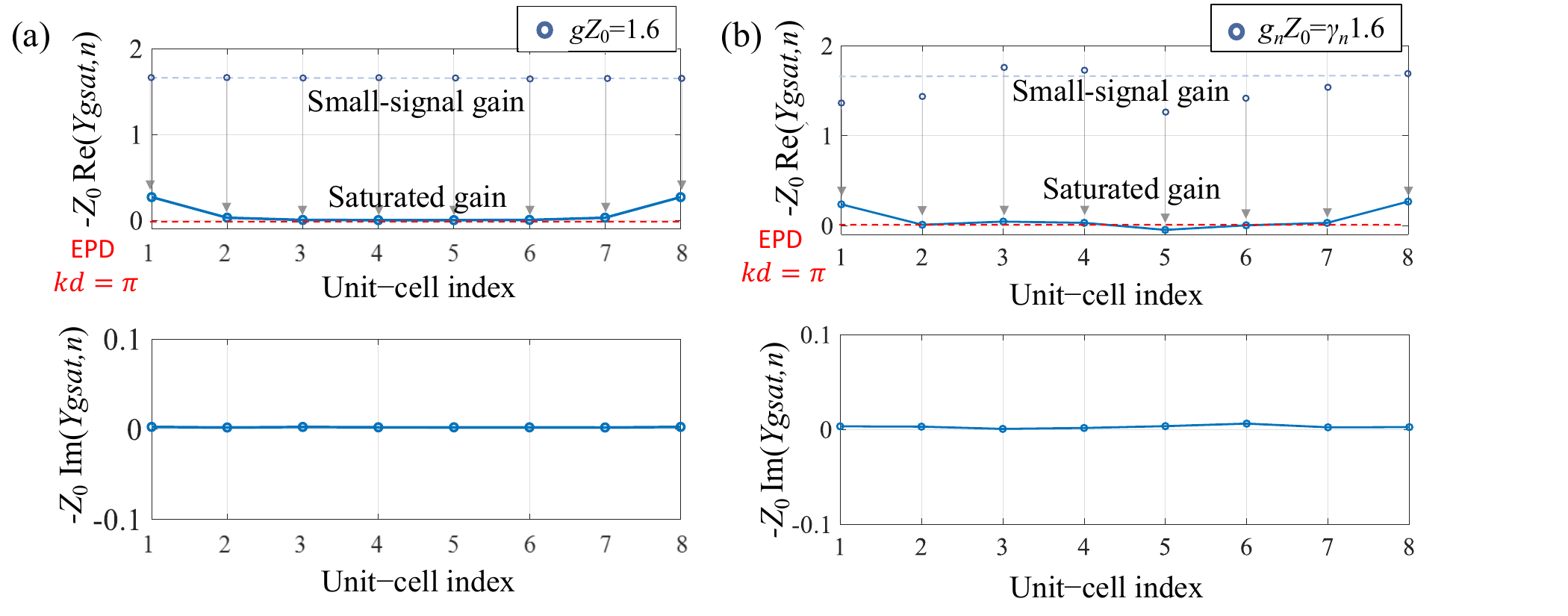}
			\par\end{centering}
		\caption{Saturated gain admittance $-Y_{gsat,n}$ calculated at each unit cell (blue curve), for an array with $N=8$ unit cells. The radiating elements have conductance $Y_{r}Z_{0}=2.5$. We consider two cases: (a) all unit cells have the same small-signal gain $gZ_0=1.6$; and (b) each unit cell has a different small-signal gain  $g_nZ_0=\gamma_n 1.6$, with factors $\gamma_{n}$ equal to $85\%$, $90\%$, $110\%$, $105\%$, $82\%$, $89\%$, $95\%$, $101\%$, respectively, of the EPD small-signal gain. The saturated gain $-Y_{gsat,n}$  (blue) converged to a much smaller value than the initial small-signal gain, and tends to vanish for longer arrays. For infinite arrays it converges to $Y_{gsat,n}Z_0=0$ that is associated to an EPD at 3 GHz with $kd=\pi$, shown with a dashed red line.\label{fig:Gain_variation_EPD}}
	\end{figure*}

	\begin{equation}
		{\boldsymbol{\Psi}}(z+d)=\underline{\mathbf{T}}_{\textrm{U}}{\boldsymbol{\Psi}}(z),
	\end{equation}
 
\noindent where $\underline{\mathbf{T}}_{\textrm{U}}$ is the $2\times2$ transfer matrix relative to a unit cell denoted by a dashed line in Fig. \ref{fig:Waveguide_nolinear}(a). It is built using the sub blocks relative to shunt gain $\underline{\mathbf{T}}_{\textrm{gain}}$, shunt loss $\underline{\mathbf{T}}_{\textrm{loss}}$, and two lossless transmission lines with electric length $\theta=k_{w}d/2$, $\underline{\mathbf{T}}_{\textrm{TL}}$. Multiplying each segment's transfer matrix yields the unit-cell's transfer matrix
	
	\begin{equation}
		\underline{\mathbf{T}}_{\textrm{U}}=\underline{\mathbf{T}}_{\textrm{gain}}\underline{\mathbf{T}}_{\textrm{TL}}(\theta)\underline{\mathbf{T}}_{\textrm{loss}}\underline{\mathbf{T}}_{\textrm{TL}}(\theta).
	\end{equation}
	
	 The eigenmodes supported by the waveguide system are found by solving the eigenvalue problem 
	\begin{equation}
		\left(\underline{\mathbf{T}}_{\textrm{U}}-\lambda\underline{\mathbf{I}}\right)\boldsymbol{\Psi}(z)=0,\label{eq:Eigenvalue prob}
	\end{equation}
	in which $\underline{\mathbf{I}}$ is the identity matrix of order two \cite{Abdelshafy2021Exceptional}. Eigenvalues are in the form of $\lambda_{i}=e^{-jk_{i}d}$, with $i=1,2$, and $k_{i}$ is the Floquet\textendash Bloch modal wavenumber in the periodic waveguide.

	The eigenvalues are obtained by calculating the roots of the characteristic equation $\det(\underline{\mathbf{T}}_{\textrm{U}}-\lambda\underline{\mathbf{I}})=0$ that is
	
	\begin{equation}
		\begin{array}{c}
			\lambda^{2}+\Big[-2\cos(2\theta)+Y_{g}Y_{r}Z_{0}^{2}\sin^{2}(\theta)\ \ \ \ \ \ \ \ \\
			\ \ \ \ \ \ \ \ \ \ \ \ \ -jZ_{0}Y_{r}(1+Y_{g}/Y_{r})\sin(2\theta)\Big]\lambda+1=0.
   
		\end{array}\label{eq:Char_Eq}
	\end{equation}

 \begin{figure*}[t]
		\begin{centering}
			\includegraphics[width=7in]{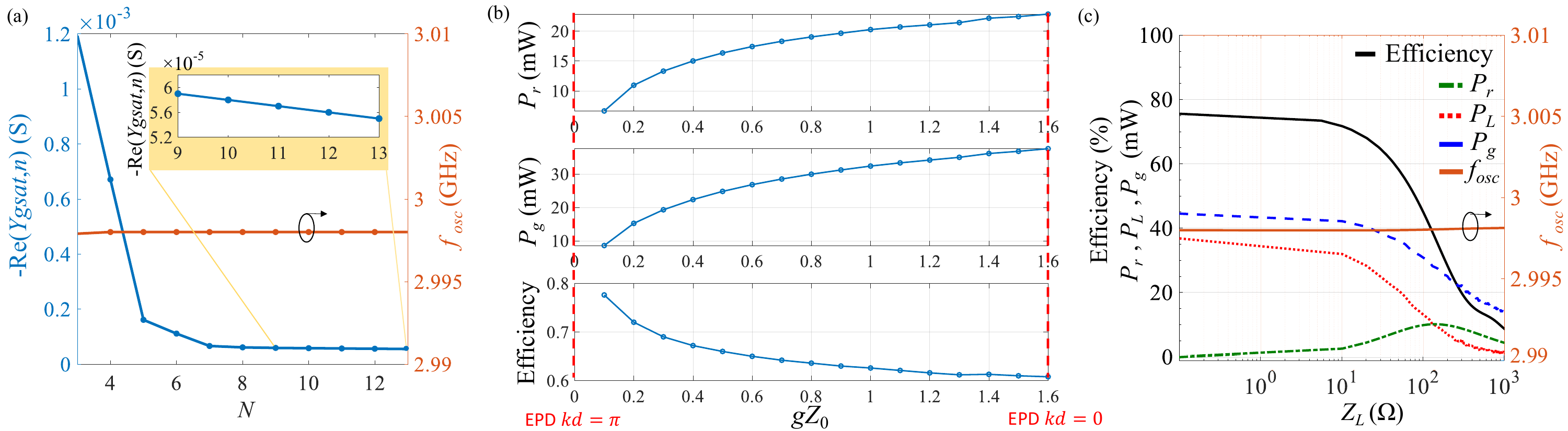}
			\par\end{centering}
		\caption{(a) Saturated nonlinear gain at the middle of the structure ($-Y_{gsat,n}$) (blue) and oscillation frequency (orange) versus the total number of unit cells $N$.(b) Radiated power $P_{r}$ delivered to the periodic elements with conductance $Y_{r}$. The total power $P_g$ is delivered by the nonlinear gain elements $Y_{gsat,n}$. The efficiency is $P_{r}/P_{g}$. The array has $N=8$ unit cells and $Z_{0}=50\:\Omega$. The small-signal is $gZ_{0}=1.6$, and the radiation conductances have $Y_rZ_{0}=2.5$.  (c) Delivered power to the radiating elements $P_r$, to the loads on the left and right $P_L$, and delivered by the nonlinear gain elements $P_g$, versus load variation $Z_{L}$. We also show the efficiency $P_r/P_g$ and the oscillation frequency $f_{osc}$.  \label{fig:Gain_variation-1}}
	\end{figure*}
 
 The characteristic equation is a second-order polynomial that results in $k_{2}=-k_{1}$. In the Eq. \ref{eq:Char_Eq}, the $\lambda$'s coefficient equal to $\pm2$ is a sufficient and necessary condition to have a second order degeneracy of the eigenvalues at $\lambda_{e}$, leading to the two EPD conditions
	
	\begin{equation}
		\begin{array}{c}
			-2\cos(2\theta)+Y_{g}Y_{r}Z_{0}^{2}\sin^{2}(\theta)=\pm2\\
			jZ_{0}Y_{r}(1+Y_{g}/Y_{r})\sin(2\theta)=0,
		\end{array}.\label{eq:EPD_condition}
	\end{equation}
where $+2$ corresponds to an EPD at $kd=\pi$ and the $-2$ corresponds to an EPD at $kd=0$. The degeneracy conditions are discussed in \cite{Abdelshafy2021Exceptional} with more details. Here, we just provide a brief summary of the conditions to obtain the EPD for the special case with $\theta=\pi/2$ at $3$ GHz where the phase velocity to unit-cell period ratio is  $v_{w}/d$= $6\times10^{9}\:\mathrm{s}^{-1}$. There are three possible degeneracy conditions with real $Y_g=-g$: (i) no linear gain or loss ($Y_{r}=0$ or $g=0$); (ii) Symmetric linear gain and loss ($Y_{r}Z_{0}=gZ_{0}=2$) results in an EPD at $kd=\pi$; (iii) Asymmetric linear gain and loss where the EPD at $kd=0$ happens when $g Z_{0}=4/(Y_{r}Z_{0})$ and the EPD at $kd=\pi$ happens when $g=0$. In this paper, we focus on the EPD condition (iii), i.e., the asymmetric linear gain and loss case, and analyze the nonlinear effects. To confirm the coalescence of the eigenvectors in our system, we use the concept of coalescence parameter $C$ (also called hyperdistance) \cite{Nada2017Theory, Abdelshafy2019Exceptional}. The coalescence parameter is a mechanism to measure the separation between the eigenvectors and how close they are to their degeneracy. The coalescence parameter vanishes when the eigenvectors collide. Thus, the Hermitian angle $\phi$ between the eigenvectors $\boldsymbol{\Psi}_{1}$ and $\boldsymbol{\Psi}_{2}$ is defined as \cite{Scharnhorst2001Angles,Galantai2006Jordans}.
	
	\begin{equation}
		C=\left|\sin\left(\phi\right)\right|,\:\:\: \:\:\:  \cos\left(\phi\right)=\frac{\left|\left\langle \boldsymbol{\Psi}_{1},\boldsymbol{\Psi}_{2}\right\rangle \right|}{\left\Vert \boldsymbol{\Psi}_{1}\right\Vert \left\Vert \boldsymbol{\Psi}_{2}\right\Vert }.
	\end{equation} 
  
  The $\cos\left(\phi\right)$ is found by using the inner product $\left\langle \,\right\rangle $,
	absolute value $\left|\,\right|$ and norm of a complex vector $\left\Vert \boldsymbol{\Psi}\right\Vert =\sqrt{\left\langle \boldsymbol{\Psi},\boldsymbol{\Psi}\right\rangle }$. 
	When the $\sin\left(\phi\right)=0$ two eigenvector coalesce at the eigenvectors $\boldsymbol{\Psi}_{e}$ defining the EPD corresponding to the eigenvalue $\lambda_{e}$. Figure. \ref{fig:2nd_EPD} (b) shows the coalescence parameter for different normalized gain values with a fixed loss $Y_{r}Z_{0}=2.5$, showing two degeneracies of the eigenvectors at $kd=\pi$ (when $gZ_{0}=0$) and $kd=0$ (when $gZ_{0}=1.6$). Figure \ref{fig:2nd_EPD}(c) shows the dispersion relation of complex-valued wavenumber versus frequency for the case with $Y_{r}Z_{0}=2.5$ and $gZ_{0}=1.6$, where the EPD happens at $f=3\:\mathrm{GHz}$ at $kd=0$. The dispersion is fitted by the quadratic curve $(f-f_{e})\propto\eta(k-k_{e})^{2}$, where $f_{e}$ is the frequency at which the two modes coalesce, and $k_{e}$ is the wavenumber at the degeneracy point (black dashed line). The flatness coefficient $\eta$	($\mathrm{m/s^{2}}$) shows the flatness of the dispersion in proximity of the degeneracy and it is related to $\partial^{2}f/\partial k^{2}$. A lower value of $\eta$ means a flatter dispersion, and by engineering the structure,  the desired parameters could be achieved. In the studied case we have  $\eta\approx2.06\times10^{5}\:\mathrm{m^{2}/s}$. 
 
 Note that Figs. \ref{fig:2nd_EPD}(b) and (d) exhibit another EPD at $f=3\:\mathrm{GHz}$ and $kd=\pi.$ The EPD is found for loss and gain at $Y_{r}Z_{0}=2.5$ and $gZ_{0}=0$, respectively. The dispersion is fitted by the quadratic formula	$(f-f_{e})=\pm\eta(k-k_{e})^{2}$ with $\eta\approx7.153\times10^{4}\:\mathrm{m^{2}/s}$, denoted by the black dashed line. As a reference, we also show in Fig. \ref{fig:2nd_EPD}(e) a dispersion relation for a case without EPD with gain $gZ_{0}=0.8$ and loss $Y_{r}Z_{0}=2.5$. We see that the eigenvalues are crossing but we do not have a degeneracy and the relation $(f-f_{e})\propto\eta(k-k_{e})^{2}$ is not satisfied.

	\section{Oscillator Operating at Second Order EPD\label{sec:Oscillator}}

	We now consider the waveguide with EPD at $kd=0$ at $3$ GHz, by selecting small-signal gain $gZ_{0}=1.6$ and radiation loss $Y_{r}Z_{0}=2.5$, while each segment has the same electric length as previously, with $\theta=\pi/2$ at $3$ GHz. Then, we set the number of unit cells to be $N=8$, and when using nonlinear gain as in Eq. (\ref{eq:Nonlinear_gain}) and we observe that the system  oscillates at $f_{osc}=3\:\mathrm{GHz}$. Since the system was designed at the EPD with $kd=0$ (based on the small-signal gain value), one would expect the signal of each unit cell to have the same phase of oscillation. But, in reality,  we observe that contiguous unit cells have a $\pi$ phase difference. In other words, under the small-signal condition, the system should operate at the EPD with $kd=0$, but we observe that after reaching saturation it operates at another point.
 
  To gain physical insight, we calculate the saturated gain $Y_{gsat,n}$ on each unit cell as shown in Fig. \ref{fig:Gain_variation_EPD} (a). For example, the saturated gain of the 4th active element of the array was found to be $-Y_{gsat,4}Z_{0}=0.003$, which is very far away from the starting small-signal one ($g_nZ_{0}=1.6$). The other saturated gains $Y_{gsat,n}$ follow an analogous behavior, as shown in Fig. \ref{fig:Gain_variation_EPD} (a).
  We then look at the color map in Fig.  \ref{fig:2nd_EPD}(b) and observe that there is another point where the coalescence parameter $C$ vanishes, at $gZ_0=0$, yielding a different  EPD condition. From Fig. \ref{fig:2nd_EPD}(d), we also observe that the phase shift associated to this EPD has $kd=\pi$, which is what we are observing in the saturated regime, even if the system started from the other EPD condition associated with $kd=0$.   
  
In summary, results from an array with a finite number $N$ of elements indicate that the system tends to work at the point where the real and imaginary parts of the saturated gain is close to zero (as verified next). From the dispersion diagram of the complex-valued wavenumber $k$ versus frequency, we have verified that the saturation point is at a gain value that corresponds to another EPD condition with $kd=\pi$, which explains the observed $\pi$ phase shift in the time-domain waveform.

  As further confirmation of these interesting dynamic properties, we also investigate the case where the nonlinear small-signal gain in each unit cell is {\em nonuniform}, i.e., varies as $g_{n}Z_{0}=\gamma_{n}1.6$ from the EPD value $g_{n}Z_{0}=1.6$. For the case shown in Fig. \ref{fig:Gain_variation_EPD} (b), small-signal gain varies with an arbitrary value $-15\%<\gamma_{n}<15\%$. Specifically for the mentioned arbitrary case, nonlinear active elements with small-signal gains are set on each unit cell with $\gamma_{n}$ as $85\%$, $90\%$, $110\%$, $105\%$, $82\%$, $89\%$, $95\%$, $101\%$, respectively. After the system reaches saturation, the time domain waveforms show that the system tends again to work at the point where the saturated gain in each unit cell tends to be uniform and such that $-Y_{gsat,n} Z_0<<1 $. The saturated gain values in Fig. \ref{fig:Gain_variation_EPD} (b) are very close to those in Fig. \ref{fig:Gain_variation_EPD} (a). Thus, random variations in small-signal gain in each unit cell do not affect the system's saturation regime at the EPD.

We have performed time domain calculations on arrays with various lengths $N$ and observed how the length affects the saturation regime. We considered nonlinear active elements with small-signal gain of $gZ_0=2$. Figure \ref{fig:Gain_variation-1} (a), shows the saturated gain in the middle of the array $-Y_{gsat,n}$ in the middle of the array (where $n$ is either $(N+1)/2$ or $N/2$ for odd or even $N$, respectively) for different array lengths $N$ (blue line), and the corresponding oscillation frequency  after reaching the saturated regime (orange line). As $N$ increases, the saturated gain has a monotonic decrease, as shown in the inset of Fig. \ref{fig:Gain_variation-1} (a), indicating that the saturated system seems to converge to the EPD at $kd=\pi$ that occurs at 3 GHz. Indeed, in the saturation regime, the array  exhibits stable oscillations at $f_{osc}=3\:\mathrm{GHz}$.

We then investigate the efficiency of the system in terms of radiating power with respect to the power arising from the gain elements for the case of an array with $N=8$ gain elements. In the saturated regime, we calculated the power $P_r$  radiated by the radiating elements, and the one delivered by the nonlinear gain elements $P_g$. The efficiency of the system is defined as the ratio $P_r/P_g$ evaluated in the saturated regime. Simulations are performed for different values of uniform nonlinear small-signal gains, ranging from $gZ_{0}\approx0$ to $gZ_{0}=1.6$, as shown in Fig. \ref{fig:Gain_variation-1}(b). These two values represent the two small-signal gains that are associated to the two EPDs in Fig. \ref{fig:2nd_EPD}(b).  The results reveal that when the system starts from a small-signal gain close to zero,  after reaching saturation it operates with maximum efficiency. The efficiency decreases when the system starts from a larger value of small-signal gain; the efficiency is low even in the case when the system starts from the EPD associated with $gZ_{0}=1.6$ and $kd=0$. Overall, the oscillating array system offers stable frequency of oscillation for all small-signal gain values, and high efficiency in radiating power for the lower end of small-signal gain values.

  Additionally, we investigate the impact of the two load impedances $Z_L$ on the saturation regime, displayed in Fig. \ref{fig:Gain_variation-1}(c). Time-domain simulations have been performed for $Z_{L}$ ranging from 0.1 $\Omega$ to 1 $\mathrm{k}\Omega$, and then the power in the saturated regime has been evaluated. We still assume $N=8$ nonlinear gain elements  with $gZ_{0}=1.6$ and radiating elements with  $Y_rZ_{0}=2.5$.  Figure \ref{fig:Gain_variation-1}(c) shows the power $P_r$ radiated by the arrayed radiating elements  (dashed green), the power $P_L$ delivered to the two load terminations  (dashed red), and the power delivered by the nonlinear gain elements in saturation (solid blue).
   Additionally, the figure presents the efficiency of the structure defined as $P_r/P_g$ (solid black), and the oscillation frequency (solid orange). These outcomes demonstrate that remarkably the oscillation frequency at $f_{osc}=3\:\mathrm{GHz}$ remains almost constant, i.e., almost equal to the EPD one, with a negligible shift of only 0.006\% (\ensuremath{\sim}200 KHz) when varying $Z_L$ from $0.1\:\Omega$ to $1\:\mathrm{k\Omega}$. The stability of the oscillation frequency over a wide range of variations of the load resistance and also of the array length (number of unit cells) shows the robustness of the proposed array oscillator whose saturated regime always converges to an EPD state. 
  
\section{Tunability}

 \begin{figure}[t]
		\begin{centering}
			\includegraphics[width=2.9in]{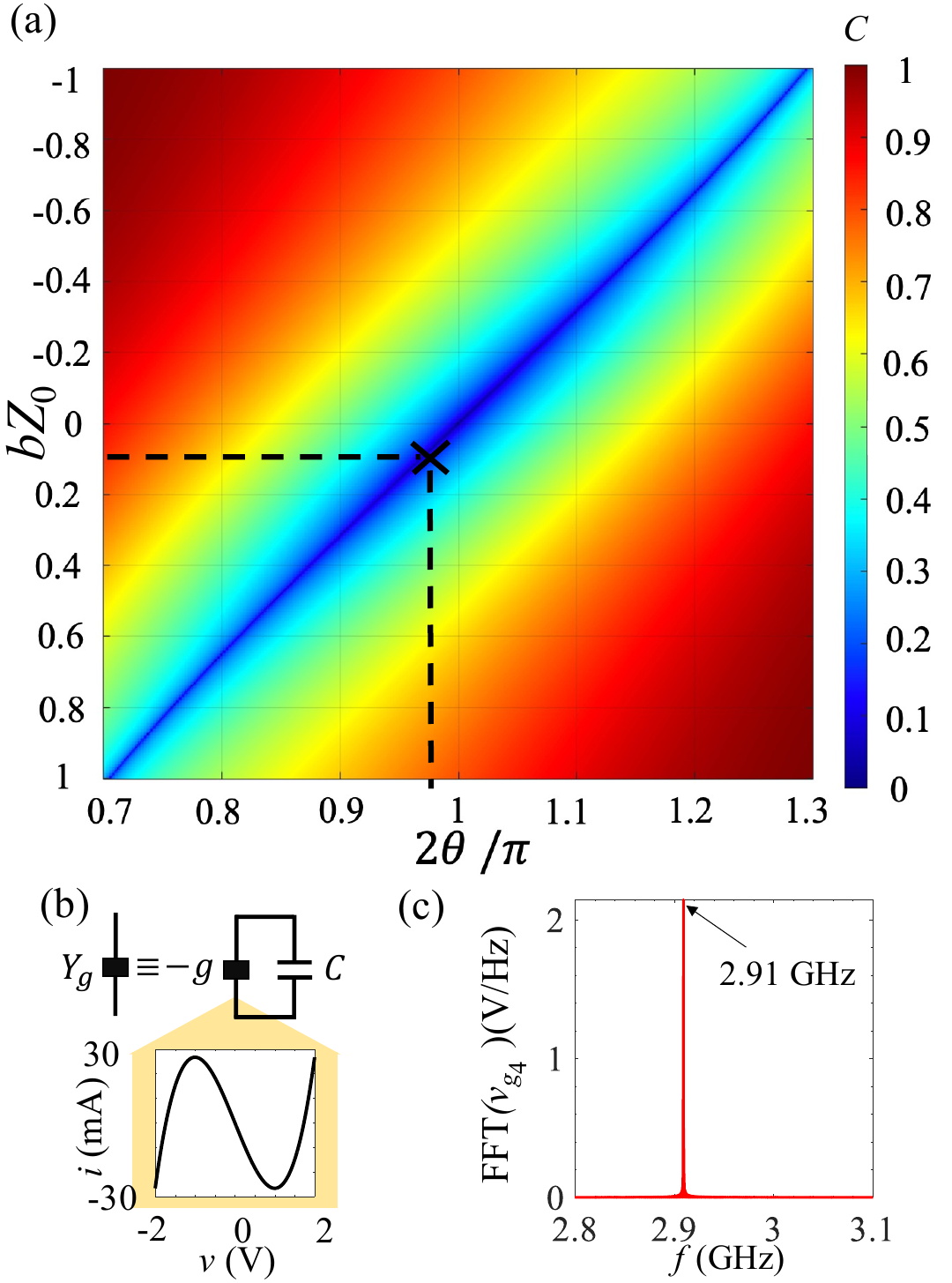}
			\par\end{centering}
		\caption{(a) The vanishing of the coalescence parameter shows different values for reactance satisfies the EPD condition. The EPs happen in different $\theta$ for different reactive susceptance $b$. (b) The frequency spectrum of the nonlinear gain voltage $v_{4}(t)$, which shows the oscillation frequency shifts when we added a small reactance to all nonlinear gain through the structure at $f_{osc}=2.91\:\mathrm{GHz}$  \label{fig:tunability}}
	\end{figure}

 The effect of the reactive part in the active element components $Y_g$ is analyzed here. Such reactance can be present because of parasitic effects (e.g., in the solid state device or in its  packaging, because of the way they are mounted, etc.) or because it can be added for tuning purposes. Therefore, even for small signals, we assume that the uniform linear gain elements $Y_g$ are not a purely real negative conductance and we consider the effect of an additional inductive/capacitive susceptance as $Y_g=-g+jb$. The EPD condition corresponding to $kd=\pi$ (i.e, the $\lambda$ coefficient in Eq. (\ref{eq:Char_Eq}) is equal to $+2$) is discussed in the following. Accounting for the reactive part $b$, the EPD condition leads to these two new equations by setting  both the imaginary and real parts equal to zero: 
	
	\begin{equation}
		\begin{array}{c}
			bY_{r}\sin^{2}(\theta)+Y_{r}(1-g/Y_{r})\sin(2\theta)=0\\
			\\
			-Y_{r}g\mathrm{\sin^{2}(\theta)}-b\sin(2\theta)-4\cos^{2}(\theta)=0
		\end{array}\label{eq: EPD condition}
	\end{equation}
	
	Therefore, by setting  $g=0$ and $b=-2\cos \theta/\sin\theta$, the degeneracy condition at $kd=\pi$ is satisfied. Hence, we can still find an EPD if the active reactance value is $b=-2\cos \theta/\sin\theta$, which implies that $\theta \ne \pi/2$ anymore, hence the EPD frequency is not at 3 GHz anymore. This analysis also clarifies why the results of Sec. \ref{sec:Oscillator}, based on assuming that $\theta=\pi/2$ at $3$ GHz, implied that the imaginary part of the saturated nonlinear gain approached zero.

 Figure \ref{fig:tunability}(a) shows the vanishing of the coalescence parameter $C$ by varying $x$ and $\theta= k_w d/2$, where $k_w=2\pi f/v_w$, assuming $\mathrm{Re}(Y_g)=-g=0$. Since the vanishing of the coalescence parameter indicates the occurrence of an EPD, the results illustrate how EPDs occur at  frequencies $f=\theta v_w/(\pi d)$ that depend on $b$. The EPD frequency decreases/increases for larger capacitive/inductive $b$. For instance, we assumed that the uniform nonlinear small-signal gain admittance $Y_g=-g+jb$ comprises a capacitive reactive susceptance  $b=j2\pi f C$  ($C= 0.1$ pF) equal to $bZ_0=0.091$ at $2.91$ GHZ. 
 
We then perform the nonlinear time-domain simulation for an array of $N=8$ gain elements, assuming a small capacitor of $0.1\:\mathrm{pF}$ in parallel to the nonlinear small-signal gain conductance $g$ in each unit cell as depicted in Fig. \ref{fig:tunability}(b). Time domain results obtained by the Keysight ADS simulator show that after saturation the oscillation frequency is $f_{osc}=2.91\:\mathrm{GHz}$, corresponding to the EPD point denoted by the cross symbol in Fig. \ref{fig:tunability}(a). Analogously to what has been demonstrated in the last section, here we start from a given value of small-signal gain, and after reaching saturation the admittance is  $Y_{gsat,n}Z_{0}\approx 0+jb$. For example, assuming the  nonlinear small-signal gain conductance of $gZ_0=0.3$, the saturation gain on 4$th$ unit cell calculated at the oscillation frequency $f_{osc}=2.91\:\mathrm{GHz}$ is $Y_{gsat,4}Z_0 =-0.009+j 0.093$ where the reactive part mostly comes from the added capacitor. Thus, the numerical result confirms that the system once again after starting from a given small-signal admittance value (including a reactance) converges to the limiting value associated to the EPD at $kd=\pi$ where the saturated admittance is  $Y_{gsat,n}=0+jb$. In summary, an additional reactive part of the nonlinear gain can be adjusted to achieve tunability of the frequency of the stable oscillation.

	\section{Conclusions}
	
	We have investigated the effect of nonlinear active elements in a waveguide with a periodic array of radiating elements and discrete gain elements that supports EPDs. Importantly, by using a nonlinear gain for each active element in a finite-length array, we have demonstrated that for arbitrary choices of gain values, the array reaches a stable oscillation regime operating at a specific EPD at $kd=\pi$, independently of the number of array elements. Our results showed that even a $15\%$ arbitrary variation in small-signal gain in each of the nonlinear elements does not alter the overall performance and the saturated gain is uniform and independent of the initial choice of the small-signal gain because the saturated gain is associated to an EPD. In other words, we have demonstrated the concept that the saturated system tends to operate at the exceptionally degenerate eigenmode. However, the initial choice of nonlinear small-signal gain affects the radiation power efficiency.
 We have also demonstrated that not all EPD are desirable points of operation, indeed in one representative example the initial choice of small-signal gain was an EPD but the system migrated to another EPD after reaching saturation. Further studies are needed to provide a more comprehensive analysis of nonlinear dynamics of systems with EPDs.
 
  The proposed strategy to conceive coherent arrays of oscillators based on EPD operation provides a stable oscillation frequency for efficiently extracting radiation power (even when there is an unbalanced nonlinear gain across the array). Furthermore, the oscillation frequency does not change when varying the array length and the impedance loading at the two array ends. Moreover, we have shown also the EPD condition in the presence of reactance in the active elements and justify the EPD occurrence by using the vanishing of the coalescence parameter. Our analysis revealed that the EPD occurs at a frequency that depends on the shunt capacitive or inductive reactance values. Nonlinear time-domain simulation has shown that a tunable EPD oscillator with a stable oscillation frequency is conceived by adding small shunt capacitances to all the gain elements. 
  
  This proposed EPD oscillating array scheme may have diverse applications, including radiating arrays of active integrated antennas, and distributed high-power oscillators.  The fundamental principles analyzed in this paper are valid for a synchronized array of oscillators, from radio frequency to optics. The demonstrated concept is applicable also to high-power lasers with distributed power extraction (like in a vertically emitting laser). The proposed strategy is  general, and it focuses on demonstrating possible advantages of EPD-based distributed oscillators in the robustness of oscillation, coherence of radiation with small phase noise, and high power in large apertures of radiators.

	\section*{Acknowlegment}
	
	This material is based upon work supported by the USA National Science Foundation under Award NSF ECCS-1711975.
	
	\appendix{}


\end{document}